\documentclass[aip, amsmath,amssymb,
 reprint,%
]{revtex4-1}
\usepackage{graphicx}
\usepackage{dcolumn}
\usepackage{bm}

\usepackage{epsfig}
\usepackage{amsmath}
\usepackage{makecell}
\usepackage{multirow}
\usepackage{blindtext}
\usepackage{hyperref}
\usepackage{amssymb}
\usepackage{ragged2e}
\usepackage{threeparttable,booktabs}
\usepackage{diagbox}
\usepackage{hhline}
\usepackage{color}
\usepackage{verbatim}
\usepackage{adjustbox}

\usepackage[utf8]{inputenc}
\usepackage[T1]{fontenc}
\usepackage{mathptmx}
\usepackage{etoolbox}
\makeatletter
\def\@email#1#2{%
 \endgroup
 \patchcmd{\titleblock@produce}
  {\frontmatter@RRAPformat}
  {\frontmatter@RRAPformat{\produce@RRAP{*#1\href{mailto:#2}{#2}}}\frontmatter@RRAPformat}
  {}{}
}%
\makeatother

\def\etal{{\it et al.}}

\newcommand{\RUG}{
Van Swinderen Institute for Particle Physics and Gravity,
University of Groningen, Nijenborgh 4, 9747 Groningen, The Netherlands}

\newcommand{\TAU}{
School of Chemistry, Tel Aviv University, 6997801 Tel Aviv, Israel}

\newcommand{\CU}{
Department of Physical and Theoretical Chemistry, Faculty of Natural Sciences, 
Comenius University, Mlynsk\'{a} dolina, 84215 Bratislava, Slovakia}
\usepackage{graphicx}
\usepackage{dcolumn}
\usepackage{bm}

\usepackage[utf8]{inputenc}
\begin{document}
\preprint{AIP/123-QED}
\title{Relativistic coupled cluster calculations of the electron affinity and ionization potential of Nh(113)}
\author{Yangyang Guo}
\affiliation{\RUG}

\author{Anastasia Borschevsky}
\affiliation{\RUG}
\author{Ephraim Eliav}
\affiliation{\TAU}
\author{Luk\'{a}\v{s} F. Pa\v{s}teka, *}
\email{lukas.f.pasteka@uniba.sk}
\affiliation{\CU}
\date{\today}
\begin{abstract}
 Theoretical calculations based on the Dirac--Coulomb--Breit relativistic coupled cluster method have been carried out for the electron affinities and ionization potentials of the superheavy element nihonium (Nh) and its lighter homologues In and Tl. 
 The In and Tl calculations are in agreement with measurement within uncertainties. For Nh, where experiment is yet unknown, we predict the ionization potential of 7.569(48) eV and electron affinity of 0.776(30) eV. 
 \end{abstract}
\maketitle

\section{Introduction}

All elements with atomic number greater than 104 are typically called superheavy elements (SHEs).\cite{https://doi.org/10.1002/anie.200461072} Until now the existence of long-lived stable SHEs in nature has not been verified and these elements are produced artificially
via two fusion evaporation approaches, the cold fusion or the hot fusion reactions,\cite{Moo14} at single-atom-at-a-time level. The low quantities and the short half-lifes, which decrease rapidly with increasing atomic number, make experimental spectroscopic and chemical investigations of SHEs extremely challenging.\cite{single.atom,TurPer13,sch15} These difficulties motivate theoretical studies of atomic, molecular, and even bulk properties of SHEs, both in support of the planned measurements and, often, as the only route for gaining information about electronic structure and behaviour of these elusive elements. 

The last four new elements were approved by IUPAC and added to the Periodic Table in November 2016, namely nihonium, moscovium, tennesine and oganesson.\cite{Namesandsymbolsoftheelementswithatomicnumbers113115117and118IUPACRecommendations2016} Nihonium (Nh, $Z = 113$), the first element of the heaviest 7p-block of the Periodic Table, was first synthesized in 2004  using the cold fusion reaction of lead with a bismuth target.\cite{Historyofnihonium} The volatility of Nh was investigated experimentally through its adsorption on gold and quartz surfaces,\cite{DMITRIEV2014253,TurEicYak15, Aksenov2017, YakLenDul21} but no atomic or molecular properties of this element have been measured so far. 

The aim of this work is to provide high accuracy and reliable prediction of the IP and EA of Nh. 
Ionization potentials (IPs) and electron affinities (EAs) are two of the most fundamental atomic properties. Knowledge of the IP and EA can provide insight into the chemical behaviour of an element, the type of compounds it is likely to form, and even into its solid state properties.\cite{atarah_egblewogbe_hagoss_2020}
Relativistic effects play an important role in the superheavy region of the periodic table, affecting the properties and even at times leading to change in the electronic configuration. For the 7th row elements, both the 7s and the 7p$_{1/2}$ atomic orbitals are strongly contracted and stabilized, and the spin-orbit effects lead to a large splitting between the 7p$_{1/2}$ and the 7p$_{3/2}$ orbitals.\cite{V.Pershina} These effects are expected to have a dramatic influence on the properties of Nh, compared to the other Group 13 elements.
To capture these effects we perform our calculations in a relativistic framework and use the 4-component Dirac--Coulomb coupled cluster approach with single, double, and perturbative triple excitations (DC-CCSD(T)). In order to further improve the accuracy of our predictions, we correct the DC-CCSD(T) results for the higher order effects -- the Breit and the QED contributions and excitations beyond perturbative triples. This approach leads to accuracy on the order of single meVs for properties such as IP and EA and was demonstrated to have very strong predictive power.\cite{PasEliBor17,LeiKarGuo20} Furthermore, we employ a recently developed scheme that allows us to use an extensive computational study to assign uncertainties on our predictions\cite{LeiKarGuo20,GuoPasEli21} Such uncertainties will be valuable both in future experimental research and in other theoretical studies. In parallel, we investigated the lighter homologues of Nh, In and Tl using the same approach.
For the lighter elements, we can compare our results to experiment to verify the accuracy of our calculations and to check whether the assigned error bars are realistic. By this process we provide a confirmation for our predictions for Nh. 

Previous theoretical calculations of the IP and EA of Nh were carried out using the Dirac-Fock approach\cite{osti_4176210} and later the relativistic Fock space coupled cluster method.\cite{PhysRevA.53.3926,doi:10.1021/jp8061306} Ref. \citenum{TurPer13} provides an extensive overview of the computational investigations of the molecular properties of Nh on various levels of theory. More recent studies considered the adsorption of Nh on quartz\cite{Per16} and on gold\cite{Per18} and its solid-state properties.\cite{AtaEgbHag20}

\section{COMPUTATIONAL DETAILS}
The calculations were carried out using the DIRAC19 computational program package,\cite{DIRAC19} in the framework of the relativistic 4-component Dirac--Coulomb (DC) Hamiltonian,
\begin{eqnarray}
H_\mathrm{DC}= \displaystyle\sum\limits_{i}h_\mathrm{D}(i)+\displaystyle\sum\limits_{i<j}(1/r_{ij}).
\label{eqHdcb}
\end{eqnarray}
The one-electron Dirac operator is
\begin{eqnarray}
h_\mathrm{D}(i)=c\bm{\alpha }_{i}\cdot \mathbf{p}_{i}+c^{2}\beta _{i}+V^n(i),
\label{eqHd}
\end{eqnarray}
where \( \mathcal{\alpha} \) and $\beta$ are the four-dimensional Dirac matrices and $V^n$ is the nuclear attraction operator. The finite size of the nucleus was taken into account and modeled by a Gaussian charge distribution.\cite{VisDya97}

Electron correlation was treated within the coupled cluster method at the single, double and perturbative triple excitations level (CCSD(T)). In the coupled cluster procedure, all electrons were correlated and virtual orbitals with energies above 300~a.u. were omitted.

The uncontracted all-electron correlation-consistent relativistic basis sets of Dyall of varying cardinal number ($N$) and augmentation levels (the latter representing the number of additional diffuse functions, $x$ for each angular symmetry and labeled as $x$-aug) were used.\cite{Dyall2012}
We performed extrapolation to the complete basis set (CBS) limit, using the scheme of Feller \etal\cite{Feller1992} for the DC-HF values (3-point extrapolation using 2z, 3z and 4z results) and the scheme of Helgaker\cite{HelKloKoc97} for the correlation contribution (2-point extrapolation using the 3z and the 4z results).

In order to reach meV level accuracy, one needs to address higher-order correlation and relativistic effects. For the former, the full triple and quadruple contributions were calculated with the MRCC program.\cite{doi:10.1063/1.1383290,doi:10.1063/1.2121589,doi:10.1063/1.2988052,doi:10.1063/1.5142048,doi:10.1063/1.1950567} In these calculations, the outermost $(n-1)$d and valence $n$s and $n$p electrons were correlated. Virtual orbital energy cutoff of 13 a.u. was used for CCSDT(Q) and the additional $\Delta$Q contribution was calculated with the virtual cutoff of 5 a.u. 
Dyall's 1-aug-cv$N$z basis sets were used.\cite{Dyall2012} The CCSDT results were extrapolated to the CBS limit as above, while the $\Delta$(Q) and $\Delta$Q contributions were calculated at the 3z and 2z level, respectively.
The core-valence contribution ($(n-1)$s, $(n-1)$p and $(n-2)$f shells) to the CCSDT was calculated at the 3z level. 
We have found in our earlier investigations that higher order excitations are generally localised in the valence shell region,\cite{PasEliBor17} justifying our use of a limited active space.

For the higher order relativistic corrections, we need to consider the Breit and the QED contributions. 
We correct the two-electron part of $H_\mathrm{DC}$ by the Breit transverse interaction (in the Coulomb gauge)
\begin{equation}
B_{ij} =-\bm{\alpha }_{i}\cdot \bm{\alpha }_{j}\frac{\mathrm{cos}\widetilde{\omega}r_{ij}}{r_{ij}}+(\bm{\alpha }_{i}\cdot \mathbf{\nabla}_{i})(\bm{\alpha }_{j}\cdot \mathbf{\nabla}_{j})\frac{\mathrm{cos}\widetilde{\omega}r_{ij}-1}{\widetilde{\omega}^2r_{ij}},\\
\label{eqBij}
\end{equation}
where $\widetilde{\omega}$ is the angular wavenumber of the exchange photon.

We further incorporated the QED corrections in the form of the model Lamb shift operator of Shabaev \etal\cite{ShaTupYer15} This model Hamiltonian includes the Uehling potential and an approximate Wichmann--Kroll term for the vacuum polarization potential\cite{BLOMQVIST197295} and local and nonlocal operators for the self-energy, the cross terms, and the higher-order QED terms.

The QED and the low-frequency limit ($\widetilde{\omega}\rightarrow 0$) Breit contributions
were calculated within the relativistic Fock-space coupled cluster approach (DCB-FSCCSD), using the Tel Aviv atomic computational package.\cite{TRAFS-3C}
A small correction due to the frequency-dependent part of the Breit interaction was claculated perturbatively at the mean-field DC-HF level using the GRASP92 program.\cite{DyaGraJoh89,ParFroGra96}
The calculated higher order excitation contributions and the Breit and QED corrections are added on top of the DC-CCSD(T) results.

\section{RESULTS AND DISCUSSION}

\subsection{Basis set effects}

\begin{table}[t]
  \centering
    \caption{IP and EA of In, Tl and Nh at CCSD(T) level using different quality of basis sets, in eV.}
    \begin{tabular}{@{\extracolsep{4pt}}l c c c c c c  @{}}
    \hline\hline
    \multirow{2}{*}{Basis set}&\multicolumn{3}{c}{IPs }&\multicolumn{3}{c}{EAs}\\
\cline{2-4} \cline{5-7}
     & In & Tl& Nh & In & Tl& Nh \\
   \hline
cv3z &5.695&5.975&7.337&0.152&0.131	&0.582\\
cv4z &5.758&6.071&7.486&0.241&0.230&0.689 \\ 
ae4z &5.756&6.077&7.502&0.239&0.228&0.694 \\
 1-aug-ae4z &5.759&6.079&7.509&0.304&0.277&0.718\\
  2-aug-ae4z &5.759&6.079&7.509&0.305&0.277&0.719\\
  3-aug-ae4z&5.759&6.079&--&0.305&0.277&--\\
  2-aug-aeCBS&5.804&6.148&7.613&0.315&0.299&0.777\\ 

    \hline\hline
  \end{tabular}
  \label{tab:basis set}
\end{table}

Table \ref{tab:basis set} summarizes the EAs and IPs of In, Tl and Nh for different basis set quality at the CCSD(T) level. 
Compared with cv3z, the higher cardinality cv4z basis set considerably increases all IP and EA values. This contribution has a growing trend across the IPs of the three elements ranging from 63 meV for In to 149 meV for Nh. On the other hand, for EAs we observe a mostly flat contribution of about 100 meV.
Adding the inner-core correlation functions by means of the ae4z basis set (as compared to the cv4z basis set) leads to only a small correction of at most a few meV, with the only notable exception being the increase in IP of Nh by 15 meV.
We also investigate the effect of additional diffuse functions.
The IPs and EAs differences are no larger than 1 meV between the 1-aug and 2-aug results 
and the third augmentation layer did not further improve the results. Thus, for our production calculations, we apply the CBS extrapolation scheme to the doubly-augmented all-electron 2-aug-ae$N$z basis sets. The CBS extrapolation increased the IP and EA of Nh by 104 meV and 58 meV, respectively. 
This demonstrates that the extrapolation scheme is an useful tool to improve accuracy with respect to the restricted cardinality of a finite basis set.

\subsection{Higher corrections}

\begin{table}[t]
  \centering
    \caption{IP and EA of In, Tl and Nh with higher order corrections contributions, in eV.}

  \begin{tabular}{@{\extracolsep{4pt}}l c c c c c c @{}}
    \hline\hline
    \multirow{2}{*}{Method}&\multicolumn{3}{c}{IPs }&\multicolumn{3}{c}{EAs}\\
\cline{2-4} \cline{5-7}
 & In & Tl& Nh & In & Tl& Nh \\
   \hline
  CCSD  & 5.774& 6.115 & 7.591  & 0.206& 0.206 & 0.694\\
  CCSD(T)  & 5.804& 6.148 & 7.613 & 0.315& 0.299 & 0.777 \\ 
  CCSDT    & 5.803& 6.145& 7.614& 0.343& 0.298&  0.793\\ 
  CCSDT(Q)   & 5.805& 6.145& 7.608& 0.376& 0.311&  0.777\\
  CCSDTQ   & 5.805& 6.146& 7.610& 0.374& 0.311&  0.790\\
  CCSDTQ+Breit & 5.799& 6.131& 7.567& 0.374& 0.309& 0.774\\ 
  CCSDTQ+Breit & \multirow{2}{*}{5.801}& \multirow{2}{*}{6.135}& \multirow{2}{*}{7.569}& \multirow{2}{*}{0.375}& \multirow{2}{*}{0.311}& \multirow{2}{*}{0.776}\\
  \phantom{CCSDTQ}+QED\\
    \hline\hline
  \end{tabular}
  \label{tab:corr}
\end{table}

The calculated IPs and EAs of In, Tl and Nh are shown in Table \ref{tab:corr}; the presented CCSD and CCSD(T) results were obtained using the 2-aug-ae$N$z basis sets and extrapolated to complete basis set limit. Comparing with DC-CCSD result, the perturbative triple contributions $\Delta$(T) increase the IPs of all the elements by 22--33 meV and have a larger effect on the EAs, raising them by 83--106 meV.

Both the difference between the perturbative and the full triple excitations $\Delta$T and the perturbative/full quadruple excitations $\Delta$(Q)/$\Delta$Q contribute little to the IPs of all the elements, with the largest total contribution to IP of Nh of --3 meV. 
At the same time, the higher excitations corrections are more pronounced for the electron affinities. 
The increased importance of electron correlation is to be be expected for anionic systems with 
an open valence p$^2$ configuration.
The EA of In is increased by 59 meV, while for Tl these contributions are more modest at 12 meV.
In the case of Nh, the higher excitations correction remains at 13 meV due to partial cancellation between the
the perturbative $\Delta$(Q) and full $\Delta$Q contributions. While the perturbative treatment of quadruples is apparently somewhat inadequate for Nh, the overall quadruples correction is merely --3 meV signalling convergence in terms of the excitation level.

The Breit correction decreased the IPs by 6 meV for In to 43 meV for Nh, and decreased the EA of Nh by 16 meV. The QED contributions are rather small for both properties of Nh and the lighter homologues. It is clear that Breit correction is more important for the calculation of superheavy element Nh, compared with QED correction.

\subsection{Uncertainty estimation}
\indent The extensive computational study that we carried out in this work allows us to assign reliable error bars on our calculated values, including our predictions for Nh. There are several sources of uncertainty in these calculations, stemming from the use of a finite basis set, from the incomplete treatment of electron correlation, and from the missing higher order QED corrections. We treat these sources of error separately, following the procedure outlined in Ref. \citenum{LeiKarGuo20}.

\begin{table}[t]
  \centering
    \caption{IP and EA of In, Tl and Nh obtained using different extrapolation schemes together with the resulting 95\% confidence interval, in eV.}
    \begin{tabular}{@{\extracolsep{4pt}}l c c c c c c @{}}
    \hline\hline
    &\multicolumn{3}{c}{IP}&\multicolumn{3}{c}{EA}\\
\cline{2-4} \cline{5-7}
   Scheme& In & Tl& Nh & In & Tl& Nh \\
   \hline
   Helgaker&	5.807&	6.148&	7.613&	0.312&	0.299&	0.777\\
Lesiuk&	5.819&	6.166&	7.640&	0.314&	0.304&	0.792\\
Martin&	5.797&	6.134&	7.591&	0.310&	0.294&	0.765\\
   95\% c.i.&0.021&0.031& 0.047&0.004&0.010&0.026\\ 
   \hline\hline
   \end{tabular}  
   \label{tab:scheme}
\end{table}

\begin{table}[t]
 \centering
    \caption{Main sources of uncertainty in the calculated IP and EA of In, Tl, and Nh, in meV.}
    \begin{tabular}{@{\extracolsep{5pt}}l r r r r r r@{}}
    \hline\hline
    \multirow{2}{*}{Error source} &\multicolumn{3}{c}{IP}& \multicolumn{3}{c}{EA}\\
    \cline{2-4}\cline{5-7}
    & \multicolumn{1}{c}{In} & \multicolumn{1}{c}{Tl} & \multicolumn{1}{c}{Nh}  & \multicolumn{1}{c}{In} & \multicolumn{1}{c}{Tl} & \multicolumn{1}{c}{Nh}\\
    \hline

  Basis set\\
  \quad-- CBS 	&21.4&	31.2&	47.4&	3.7&	9.7&26.3  \\ 
  \quad-- augmentation &0.0 & 0.0& 0.1 &0.0 & 0.0& 0.1  \\
  Correlation\\
  \quad-- virtual cutoff &0.5&0.5&0.6
 &0.1&0.1&0.1\\
  \quad-- higher excitations &0.9&0.3&4.3&	17.9&6.9&14.9\\
  QED  & 2.7& 4.4 & 1.8 & 1.3& 2.0 & 2.1\\
  Total&	21.5&31.5&47.6&18.3&12.1&30.3 \\ 
      \hline\hline
  \end{tabular}
    \label{tab:error}
\end{table}

The basis set contribution to the uncertainty (which often dominates the total error estimate)\cite{LeiKarGuo20} can be further divided into different components: the incompleteness of the basis set (or the error due to the choice of the CBS extrapolation scheme) and the limited number of diffuse functions.
We compare three different CBS extrapolation schemes to evaluate their effect on the calculated IPs and EAs. The Feller extrapolation scheme employed for the SCF energy has a three parameter mixed exponential form.\cite{Feller1992} To obtain the final recommended values, we combine this approach with the Helgaker extrapolation scheme for the correlation energy, which relies on the $L^{-3}$ error formula, with $L$ the highest angular momentum.\cite{HelKloKoc97}
We also used the correlation energy extrapolation scheme of Lesiuk,\cite{doi:10.1021/acs.jctc.9b00705} based on an analytic resummation of the missing energy increments using the Riemann zeta function.  
Finally, the scheme of Martin\cite{MARTIN1996669,MARTIN2006} uses the Schwartz-type extrapolation to calculate the CBS limit of both the correlation and the SCF energies. The results obtained using the different extrapolation schemes are summarised in Table \ref{tab:scheme}. We observe that the results obtained with the Helgaker extrapolation scheme fall in between the values from the other two approaches, justifying our selection of this scheme for the recommended IPs and EAs. We take 95\% confidence interval of the standard deviation between the three schemes (1.96$\sigma$) as our CBS uncertainty estimate, as shown in Table \ref{tab:error}.
We have used the 2-aug-ae$N$z basis set family in the calculations, and thus we take the difference between these and the 1-aug-ae$N$z results as an estimate of augmentation uncertainty.

Switching to the shortcomings in the treatment of relativity, we assume that the higher order QED contributions are smaller than the calculated Lamb shift and take the latter as the conservative uncertainty estimate. 

Finally, in terms of the correlation treatment, we have two sources of error: the neglect of the higher excitations beyond CCSDTQ and the finite size of the active space. The presented calculations are performed with a virtual cutoff of 300 a. We take In as a test case for investigating the effect of the cut-off and find that upon going to 5000 a.u. the calculated IPs and EAs change by only 0.007\% and 0.01\%, respectively (or less than 0.5 meV in absolute values). Using the same proportion, we estimate the cutoff uncertainty for Tl and Nh. To estimate the uncertainty due to the missing higher excitations, we take the average of the absolute $|\Delta(\text{Q})|$ and $|\Delta\text{Q}|$ contributions, to take into account the convergence behaviour between the perturbative and full excitation levels. 
 
\begin{figure}[t]
    \centering
     \includegraphics[width=\linewidth]{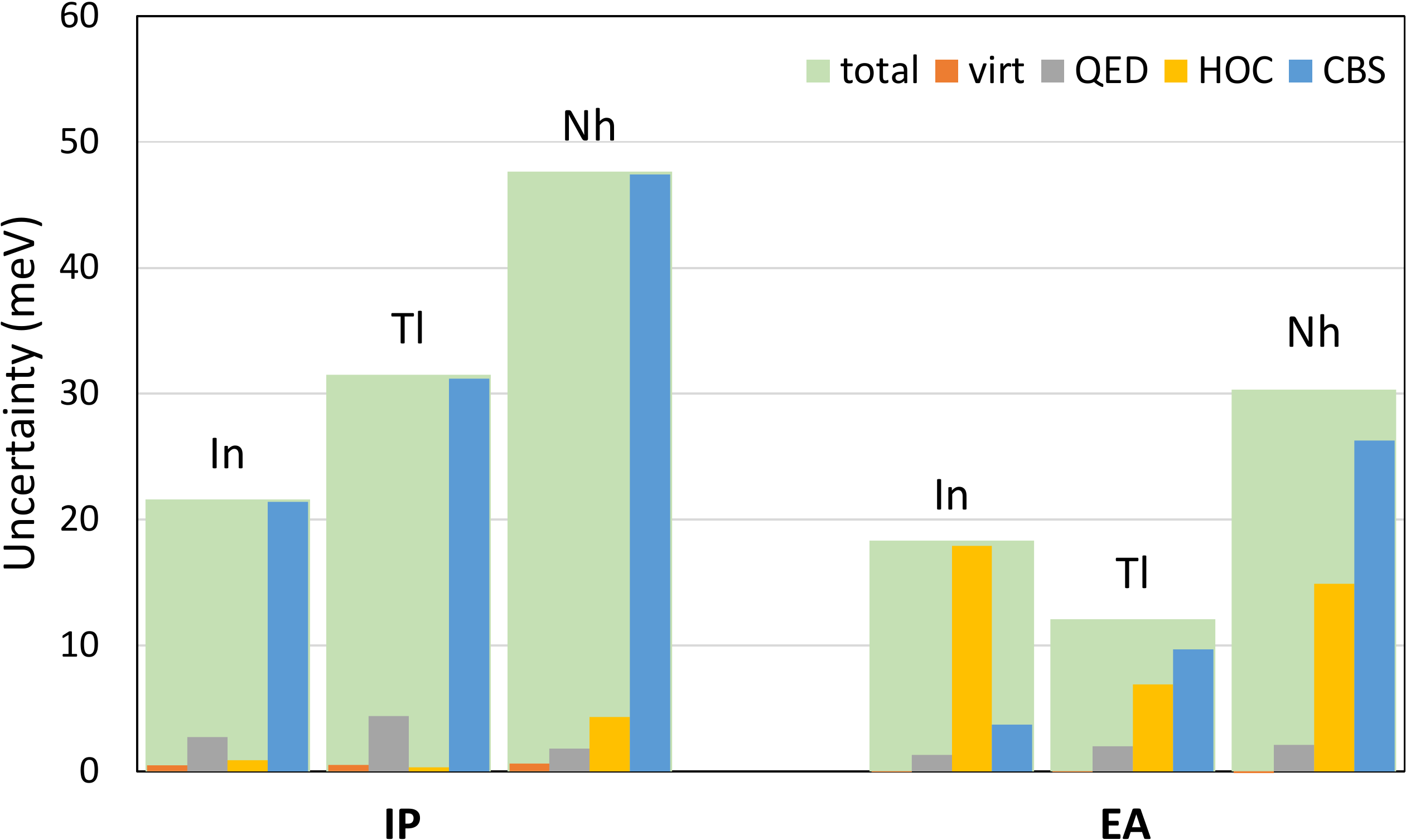}
    \caption{Contributions of the different sources of uncertainty; HOC stands for higher order correlation.}
    \label{fig:uncertainty}
\end{figure}

We assume the errors originating from the higher order effects to be independent and combine them to obtain the total uncertainty. 
Fig. \ref{fig:uncertainty} presents the relative contributions of the four sources of uncertainty, except for the negligible augmentaintion error.
In general, the dominant source of uncertainty comes from the CBS extrapolation. For both the IP and the EA, this contribution grows almost linearly with $Z$.
The higher order excitation uncertainty is more prominent for EAs compared to IPs; in the case of the EA of In, it surpasses the rather small CBS uncertainty. 

\subsection{Final results }

\begin{figure}[t]
    \centering
      \includegraphics[width=0.8\linewidth]{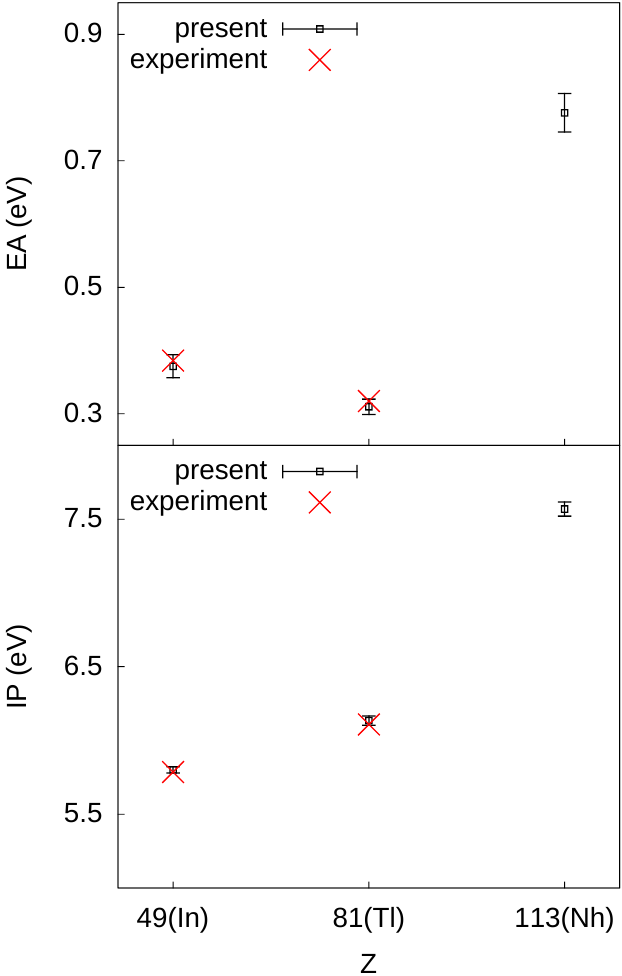}
    \caption{Calculated IPs and EAs with error bars of In, Tl and Nh (black squares), compared with experimental values for In and Tl (red crosses).}
    \label{fig:IPEA}
\end{figure}

 \begin{table*}[t]
  \centering
    \caption{IP and EA of In, Tl and Nh compared with previous calculations and experiments (eV).}
    \begin{tabular}{@{\extracolsep{4pt}}l c c c c c c c@{}}
    \hline\hline
    \multirow{2}{*}{Method} &\multicolumn{3}{c}{IP}&\multicolumn{3}{c}{EA}&\multirow{2}{*}{Ref.}\\
\cline{2-4} \cline{5-7}
 & In & Tl& Nh & In & Tl& Nh \\
\hline
   DC(B)-CCSDTQ+QED& 5.801(22)& 6.135(32)& 7.569(48)& 0.375(18)& 0.311(12)& 0.776(30)&Present\\
   DCB-MRCISD+QED& & & &  0.383(8)&0.323(7) & & \citenum{PhysRevA.104.012802}\\
   RCC-FPD & 5.786 & 6.099 & & 0.386(22)& 0.320(22)&& \citenum{doi:10.1063/1.5110174, doi:10.1063/1.5130197}\\
   DC-FSCCSD& & 6.096& 7.420& & && \citenum{doi:10.1021/jp8061306}\\
   DCB-FSCCSD  & & 6.110& 7.306& & 0.400(50)& 0.680(50)& \citenum{PhysRevA.53.3926}\\
   DC-FSCCSD  && & & & & 0.484& \citenum{PhysRevA.80.022501}\\
   DC-MCHF &  &&  & 0.398 & 0.290 & &\citenum{Li_2012} \\
   Regge pole &  &&  & 0.380 & 0.281 & &\citenum{Felfli2012a}\\
   PP-FSCCSD  & 5.780& 6.090& & 0.403& 0.347& & \citenum{doi:10.1063/1.2823053}\\
   NR-MCHF+$\Delta$RCC & &&  & 0.374(15)& & &  \citenum{Sundholm_1999}\\
   \hline
   Experiment & 5.786359(1) & 6.108194(2)&   & 0.38392(60)& 0.32005(19)&&\citenum{NEIJZEN1981271,Baig_1985,PhysRevA.82.032507,PhysRevA.101.052511}\\
   \hline\hline
   \end{tabular}  
   \label{tab:final}
\end{table*}

Our final results, including the uncertainties, are compared with a selection of previous theoretical studies conducted in the last 25 years and with experimental values for In and Tl in Table \ref{tab:final}. 
To better visualize the periodic trends, the present theoretical IPs and EAs are shown together with their corresponding uncertainties in Figure \ref{fig:IPEA} together with the experimental values.
The measured IPs and EAs of In and Tl fall within the error bars of the theoretical results in all cases, confirming the applicability of the methodology to these atomic systems.
Based on this agreement, we can have the same level of confidence in our predicted values of 7.569(48) eV and 0.776(30) eV, for the IP and EA of Nh, respectively.
The predicted IP of Nh is significantly higher than those of its lighter homologues, due to the stabilization of the 7p$_{1/2}$ orbital,  caused by the large spin-orbit splitting of the 7p shell. To illustrate, the difference of the 7p$_{1/2}$ and 7p$_{3/2}$ orbital energies reaches 115 mH at the DC-HF level; the corresponding values for In and Tl are only 10 mH and 35 mH, respectively. Similarly, we observe a much higher EA for Nh compared to In and Tl, since the relativistically stabilized 7p$_{1/2}$ orbital strongly attracts the extra electron.
These values indicate an increase in reactivity of Nh compared to its lighter homologues.
Using the predicted IP and EA of Nh, we can calculate its  electronegativity on the Mulliken scale as\cite{Mulliken1934}
\begin{eqnarray}
\chi_\mathrm{M} =\frac{\mathrm{IE}+\mathrm{EA}}{2},
\label{eqchiM}
\end{eqnarray}
resulting in a value of 4.172(39) eV which is significantly higher than the corresponding values $\chi_\mathrm{M}$(In) = 3.08514(30) eV, $\chi_\mathrm{M}$(Tl) = 3.21412(10) eV calculated from the experimental IPs and EAs.


The two most recent papers by Si \etal\cite{PhysRevA.104.012802} and Finney and Peterson\cite{doi:10.1063/1.5110174} are arguably also the most rigorous and systematic theoretical works available in the literature on the present topic. Even though both papers have investigated only the lighter homologues of Nh and not Nh itself, together with the experimental measurements, these results serve as valuable benchmarks for our methodology.
Si \etal\cite{PhysRevA.104.012802} used the DC-MCHF method with MRCISD  corrections including Breit and QED effects in a grid-based atomic code GRASP.
Finney and Peterson\cite{doi:10.1063/1.5110174} use the Feller--Peterson--Dixon (FPD) composite scheme initially based on PP-CCSD(T) with subsequent systematic corrections for the scalar and spin-orbit relativity, Gaunt and QED interactions, core-valence correlation and higher excitations. 
While our approach is different to both Si \etal\cite{PhysRevA.104.012802} and Finney and Peterson,\cite{doi:10.1063/1.5110174}
all three theoretical works agree with the experimental results available for IP and EA of In and Tl within the error margins. This offers an independent validation of our methodology.

To the best of our knowledge, the only results for IP and EA of Nh available in the literature come from the previous collaborations of the present authors.\cite{PhysRevA.53.3926,doi:10.1021/jp8061306,PhysRevA.80.022501} All three works were based on the 4-component relativistic FSCC approach. In contrast with these previous studies, the present work pushes the theory to its current computational limits, systematically accounting for all missing major contributions to IP and EA and evaluating the errors tied to the applied approximations.
The major improvement comes from addressing the basis set incompleteness in terms of cardinality and augmentation, all-electron correlation with contribuitions from triple and quadruple excitations and higher relativistic corrections (Breit, QED).
The lack of accounting for these finer contributions in the previous investigations lead to the underestimation of IP and EA of Nh.

\section{CONCLUSION\label{IV}}

We performed calculations of the IP and EA of Nh and its lighter homologues In and Tl, in the framework of the 4-component coupled cluster (CCSD(T)) method  using large uncontracted basis sets with extrapolation to the CBS limit. Breit, QED and higher order excitation corrections were added \emph{a posteriori} to the CCSD(T) values.
Deliberate analysis of error sources present in our methodology allowed us to set realistic uncertainty estimates on our results.
The calculated IPs and EAs of In and Tl match the experimental values as well as the relevant independent theoretical results, validating the employed methodology for the accurate prediction of IP and EA of Nh. 
The final predicted values for this superheavy element are 7.569(48) eV and 0.776(30) eV for IP and EA, respectively. 

\section*{Acknowledgments}

We would like to thank the Center for Information Technology of the University of Groningen for their support and for providing access to the Peregrine high performance computing cluster. LFP acknowledges the support from the Slovak Research and Development Agency (APVV-20-0098, APVV-20-0127) and the Scientific Grant Agency of the Slovak Republic (1/0777/19). Y.G. gratefully acknowledges the financial support from China
Scholarship Council. The contribution of EE was partially supported by the Ministry of Science and Higher Education of the Russian Federation within Grant No. 075-10-2020-117.

\section*{Data availability}
The data that support the findings of this study are available from the corresponding author upon reasonable request.
\bibliographystyle{apsrev4-1}
\bibliography{bib}

\end{document}